# Gearing Up for Epsilon Aurigae's First Eclipse of the Millennium

*Jeffrey L. Hopkins*
*Hopkins Phoenix Observatory*
*7812 West Clayton Drive*
*Phoenix, Arizona 85033-2439 U.S.A.*
*phxjeff@hposoft.com*

*Lothar Schanne*
*Hohlstrasse 19*
*Völklingen, Deutschland Germany*
*l.schanne@arcor.de*

*Robert E. Stencel*
*Astronomy Program - University of Denver*
*Denver, Colorado 80208 U.S.A.*
*rstencel@du.edu*

**ABSTRACT**

The mysterious 3rd magnitude long period eclipsing binary star system epsilon Aurigae is predicted to be starting its 2 year eclipse in the late summer of 2009. While this is when the real excitement starts, much is to be learned before first contact. This paper will discuss current observational results that have accumulated data using photometry, spectroscopy and other data sources. While the system is ideal for single channel photometry, due to the system brightness and distant comparison star, CCD photometry presents some interesting challenges. A fairly simple way for amateur astronomers to do BVRI CCD photometry of the system using a 50 mm camera lens and a DSI Pro camera is discussed.

## Introduction

Epsilon Aurigae is the longest known eclipsing binary star system with a period of 27.1 years. It is also a record holder for the longest eclipse which is nearly 2 years. A mid-eclipse brightening observed by several different people and methods during the last eclipse adds even more mystery. The eclipse appears to have been first noted in 1821, but was not studied in detail until the early 1900's. To add to the mystery the star system has a strange out-of-eclipse variation in brightness.

During the previous eclipse a world-wide campaign was started to observe the eclipse in detail by as many means as possible. This included both ground and spaced based observations. Very few observations were made prior to first contact or even around first contact. During 4th contact, photometric data show some wild variations in the UBV bands. For the next eclipse, first contact is predicted to be during late July or early August. From the last eclipse ingress, the contact times appear to be different at different wavelengths. An approximate 10 day delay between each UBV band was noted. The B band first contact appears to have been 12 days after the V band first contact and the U band 10 days after the B band first contact. Because of the out-of-eclipse pulsations, these times are not precise. This is an area of great interest for the next eclipse. Continuous observations are needed before and during first contact from the ultraviolet region through far infrared. These observations should be supported by continuous spectroscopy during this period. Figure 1 shows a timing schematic of the star system components from the year 2008 through 2011.

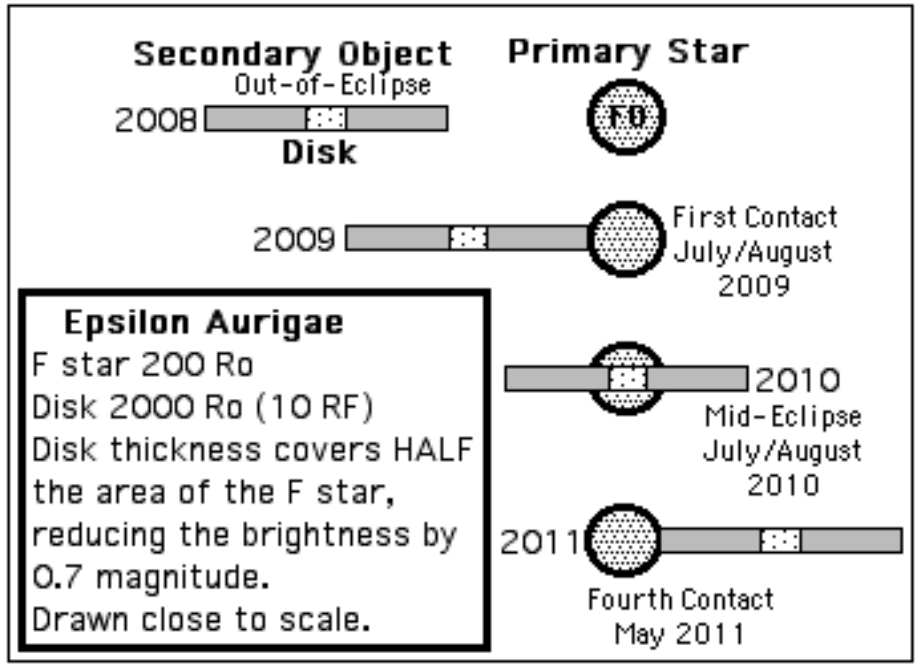

**Figure 1 Epsilon Aurigae System Timing Schematic**

**Hopkins Phoenix Observatory (HPO) Photometric Data**

One of us (Hopkins) has been actively following the star system photometrically. While data was taken at the HPO during the 1980's, no data was taken during the 1990's. Data acquisition resumed in December of 2003 and continues to date. Figure 2 shows a plot of the HPO V data from 14 December 2003 through 6 April 2008.

**Comparison Star:**
Lambda Aurigae, HR1729, HD34411
V= 4.71      B= 5.34      U= 5.46

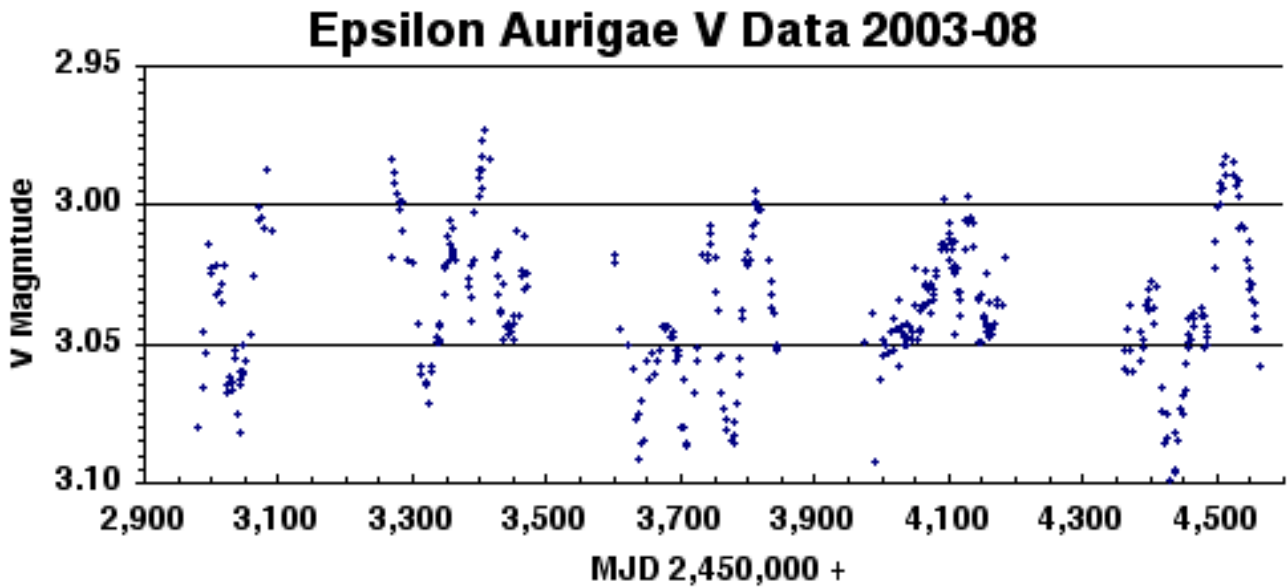

**Figure 2  HPO Data December 2003 - April 2008**

Table 1 shows a summary of the HPO data from 14 December 2003 through 6 April 2008.

| Filter | Magnitudes | | | | Freq Domain | |
|---|---|---|---|---|---|---|
| | Max | Min | Avg | Delta | Pri | Sec |
| V | 2.973 | 3.099 | 3.037 | 0.126 | 66.14 | N/A |
| B | 3.525 | 3.686 | 3.606 | 0.161 | 66.14 | N/A |
| U | 3.600 | 3.757 | 3.727 | 0.157 | 65.62 | N/A |
| (B-V) | 0.607 | 0.528 | 0.570 | 0.079 | N/A | N/A |
| (U-B) | 0.269 | 0.053 | 0.122 | 0.216 | N/A | N/A |

Table 1 HPO Data
December 2003 - April 2008

**Boyd Photometric Data**

The American Association of Variable Star Observers (AAVSO) was kind enough to give us UBV data taken by Louis Boyd using a 10" photon counting robotic telescope from September 1987 to March 2005. Table 2 shows a summary of that data.

HR1644 , HD32655  V= 6.20  B= 6.63  U= 6.93
**Note:** Boyd used a closer, but much fainter comparison star than the usually lambda Aurigae.
**Comparison Star:**

| F | Magnitudes | | | | Freq Domain | |
|---|---|---|---|---|---|---|
| | Max | Min | Avg | Delta | Pri | Sec |
| V | 2.888 | 3.111 | 2.990 | 0.223 | 97.52 | 76.89 |
| B | 3.376 | 3.844 | 3.503 | 0.468 | 97.61 | 77.02 |
| U | 3.532 | 3.907 | 3.745 | 0.389 | 97.52 | 76.92 |

Table 2 Boyd Data
September 1987 - March 2005

An effort was made to compare the HPO and Boyd data during a common period. That period was from 4 December 2003 to 27 March 2005 (JD 2,452,978 to 2,453,457 ). Table 3 shows the HPO summary data during this time and Table 4 shows the Boyd summary data during the same time frame. While the magnitudes vary between the HPO and Boyd data, the primary periods of the variations agree well at about 60 days.

| F | Magnitudes | | | | Freq Domain | |
|---|---|---|---|---|---|---|
| | Max | Min | Avg | Delta | Pri | Sec |
| V | 2.973 | 3.082 | 3.036 | 0.109 | 60.00 | N/A |
| B | 3.525 | 3.687 | 3.605 | 0.162 | 60.00 | N/A |
| U | 3.600 | 3.921 | 3.728 | 0.321 | 59.62 | N/A |

Table 3 HPO Data
December 2003 - March 2005

| Filter | Magnitudes | | | | Freq Domain | |
|---|---|---|---|---|---|---|
| | Max | Min | Avg | Delta | Pri | Sec |
| V | 2.930 | 3.105 | 2.986 | 0.175 | 59.41 | N/A |
| B | 3.427 | 3.592 | 3.508 | 0.165 | 59.47 | N/A |
| U | 3.640 | 3.897 | 3528 | 0.257 | 60.31 | N/A |

Table 4 Boyd Data
December 2003 - March 2005

**The 2007/2008 Photometric Season at HPO**

Figure 3 shows V data for the 2007 to 2008 season. Variations during the season showed a V band difference of over 0.10 magnitude with B band difference of over 0.19 magnitude and U band difference of over 0.18 magnitude. The 66 day period was consistent from what has been seen in the past, however, as the system was beginning to start dimming there was a hiccup and around JD 2,454,485 instead of decreasing the system brightened to a high of V = 2. 985, B = 3.55 and U = 3.65.

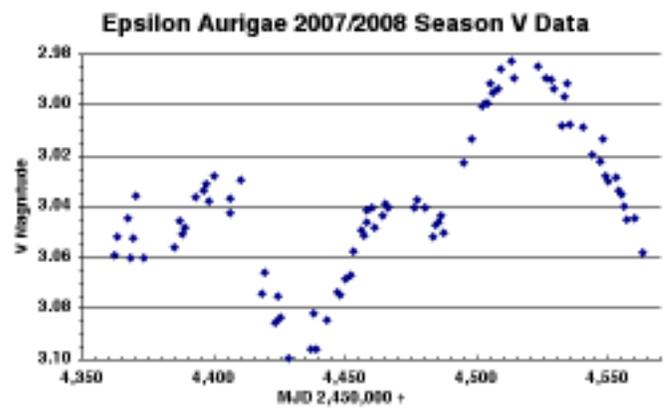

Figure 3 HPO V Data Plot
2007/2008 Season

## Frequency Domain Analysis

The frequency domain plot of the 2007 to 2008 season V filter HPO data is shown in Figure 4. The V data has a prominent period of 66.1376 days. The B band shows a prominent period of 62.42 days and the U band 64.82 days.

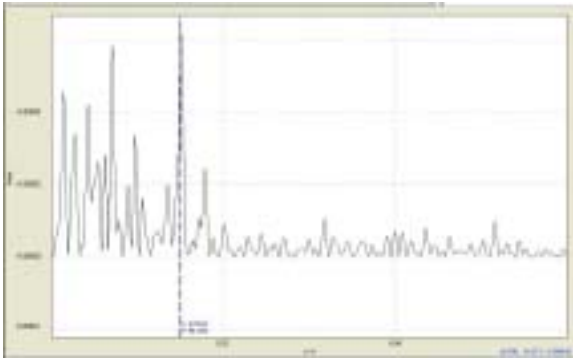

**Figure 4 HPO V Frequency Domain 2007/2008 Season**

## Photometry Methods

There are two basic methods for doing photometry, single channel and CCD. Because epsilon Aurigae is over 5 degrees from the suggested comparison star lambda Aurigae, the field of view of most telescope connected CCD systems is way too narrow. In addition the brightness of the star system causes exposure problems. While taking shorter exposures can solve the exposure saturation problem, many images must be stacked to reduce the scintillation effects. Single channel photometry including photomultiplier tube (PMT) and solid state PIN diode systems work well and avoid these problems. CCD and PIN diode systems do not work well in the ultra violet regions, however, where much interesting is going on. The PMT systems do not work well in the red, and infrared regions. However the Optec SSP-5 PMT based system can work into the red region.

## Single Channel Photometry

It is not too difficult to make your own photon counting PMT based system. The original 1P21/931A PMT based photon counting unit at the Hopkins Phoenix Observatory still provides excellent UBV photometry and has been doing so since the early 1980's. For those who prefer to buy a PMT based unit, Optec has a system for under $3,000 with filters. The SSP-5 is not a photon counting unit, however, and has limited dynamic range of 65,535 (due to a 16 bit analog to digital converter). A photon counting unit has a dynamic range greater than $10^6$ and tends to produce much higher precision than other methods, typically approaching and exceeding 0.001 magnitude in the V band.

PIN diode systems are also available from Optec. The SSP-3 allows coverage of the BVRI bands for under $2,000 with filters. The SSP-4 uses a special PIN diode that has a built-in two-stage thermoelectric cooler on the detector and allows work further into the infrared region, JH bands. The cost of the SSP-4 is around $3,000 with filters. These systems have a dynamic range of 65,535.

**Note:** For epsilon and lambda Aurigae using a SSP-4 requires a fairly large telescope. It is recommend that a telescope with an aperture greater than 10" be used with 14" or 16" ideal. With a smaller aperture the noise and drift become significantly large. With an 8" SCT the star + sky values are essentially the same as the sky values for lambda Aurigae. One nice feature about the JH band photometry is it may allow daytime observations during the interesting summer months where other photometry will not work.

Details on using the SSP-4 can be found at:
**http://www.hposoft.com/Astro/IR.html**

## CCD Photometry

As mentioned above, CCD photometry of epsilon Aurigae presents some serious problems. There are some ways around these problems, however. As Bob Buchheim[1] has suggested, an aperture stop can be used to reduce the brightness and then use a CCD system like a single channel system observing the program star and comparison star by moving the telescope. This can be a great deal of work, however. An alternate approach and one that has been experimented with at the Hopkins Phoenix Observatory uses a 50 mm F/2.0 camera lens with standard photometric BVRI filters and a Meade DSI Pro CCD camera See Figure 5. This appears to work well with repeatable results approaching 0.01 magnitudes in all bands. One secret is in defocusing the image to reduce under sampling.

Details of the 50 mm experiment can be seen at:
http://www.hposoft.com/EAur09/CCD/EAurCCD.html

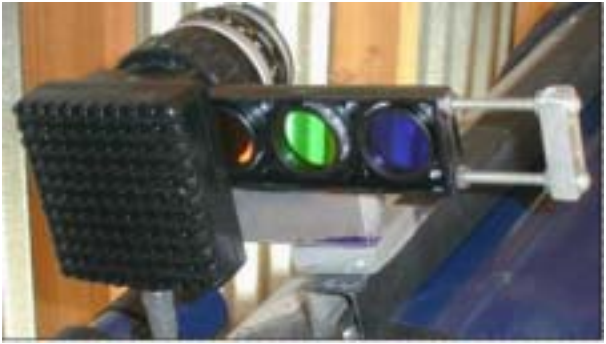

Figure 5 Wide Angle BVRI CCD Photometry

Tables 5 and 6 show a comparison of different photometry methods and what bands can be used.

| Method | U | B | V | R | I | J | H |
|---|---|---|---|---|---|---|---|
| **PMT** | | | | | | | |
| 1P21 | X | X | X | | | | |
| SSP-5 | X | X | X | X | | | |
| **PIN Diode** | | | | | | | |
| SSP-3 | | X | X | X | X | | |
| SSP-4 | | | | | | X | X |
| **CCD** | | X | X | X | X | | |

Table 5 Photometry Systems Bands

| Method | Dynamic Range | Sensitivity | QE |
|---|---|---|---|
| **PMT** | | | |
| 1P21 | > 10^6 | High | ~30 |
| SSP-5 | 65,535 | High | ~30 |
| **PIN Diode** | | | |
| SSP-3 | 65,535 | Low | ~60 |
| SSP-4 | 65,535 | Low | ~60 |
| **CCD** | ~35,000 per pixel | High | ~60 |

Table 6 Photometry Systems Characteristics

**Spectroscopy**

One of us (Schanne) has been following the star system with a spectrometer. The observatory used consists of a 100 kg self-built type mount with a Celestron C14 in a Sirius 2.4 m dome. The spectrometer is a Littrow type slit-spectrograph French product (LHIRES III, http://www.shelyak.com/) with 600, 1200 and 2400 g/mm gratings, adjustable slit with (R ≈ 3,000 to 17,000), equipped with a cooled German SIGMA 1603 ME CCD camera (http://www.nova-ccd.de) for spectrum registration and a SBIG ST-4 as slit observation camera. See Figure 6. The star in the focus of the telescope is auto guided on the slit by means of a telescope drive unit FS-2 (http://www.astro-electronic.de/) and guiding software such as MaximDL. Data reduction is done by ESO-MIDAS (http://www.eso.org/sci/data-processing/software/esomidas//). The wavelength calibration of the spectra is done by a built-in Neon calibration lamp (typical error ≈ 2 km/s).

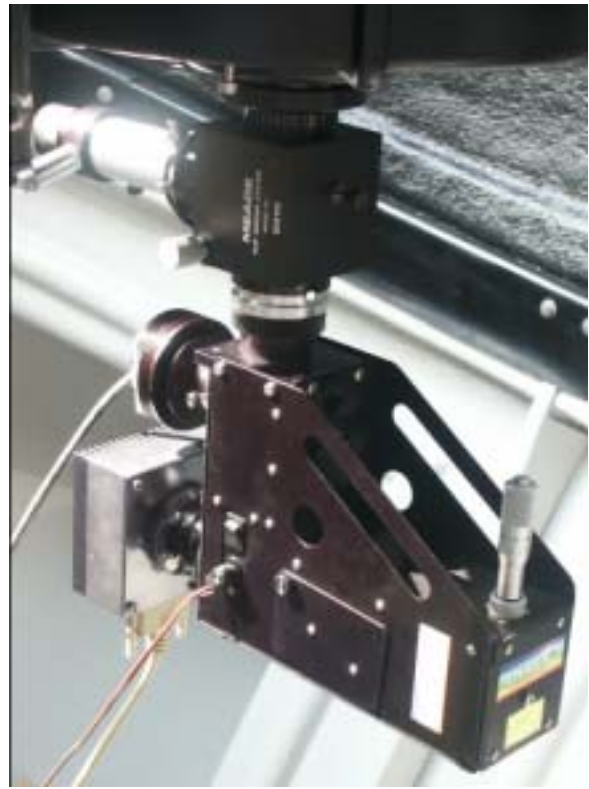

Figure 6 LHIRES III Spectrometer

Epsilon Aurigae shows typical super giant variability in the Hα line outside of the eclipse. The line profile is changing in time (LPV, line profile variation). In particular, variable emission wings on both sides of the absorption core can be seen. These LPV are seen in Figure 7.

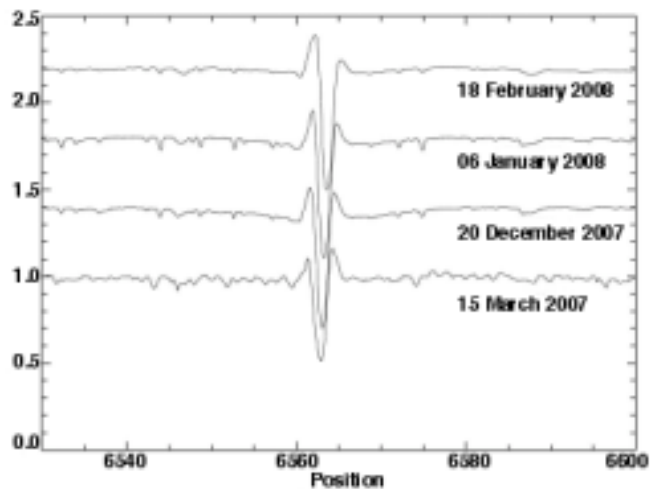

**Figure 7 Typical Line Profile Variations (LPV) of Hα line in the optical spectrum of ε Aur**

An unexpected type of LPV happened in the spring of 2005 (see Figure 8). In April and May 2005 (JD 2,453,462 to 2,453,502), the absorption component almost disappears and a fairly symmetrical pair of emission wings formed next to the line center (IBVS 5747). The exact cause of this LPV is unknown, but similar rapid changes have long been reported for this star out-of-eclipse (e.g., Wright and Kushwara, 1958)[2]. Nonetheless it is important to monitor the spectral behavior of the star system outside the eclipse to better understand the observed changes during eclipse.

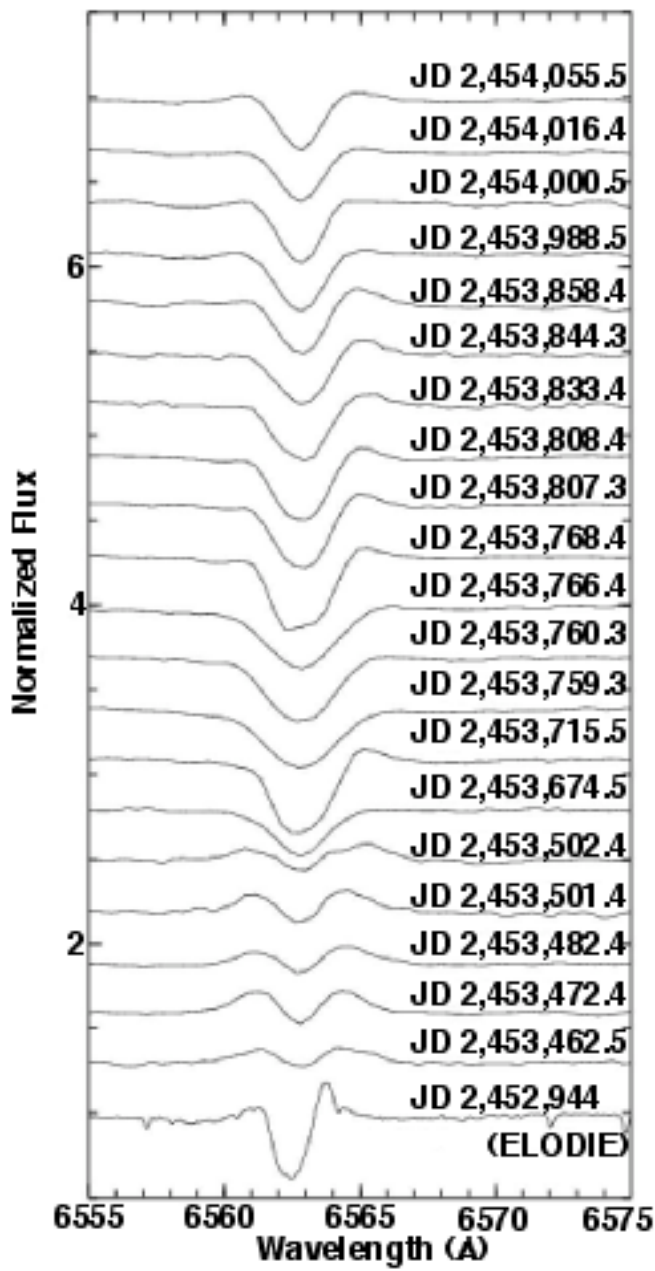

**Figure 8 Hα Line Profile Variations (LPV) of ε Aur 2005 and 2006 and reference spectrum ELODIE of November 2003.**

The strength of the emission and absorption components are characterized by the areas between the spectrum and the photospheric continuum known as the equivalent width. In Figure 9 these are shown since 2005. For comparison purposes, some HPO-photometric data (V) are included. The absorption core and emission wings show irregular change their size, but there is no simple correlation with the photometric data. Specifically, the recent 66 day period for the photometric results is not seen in the spectral behavior of the Hα line.

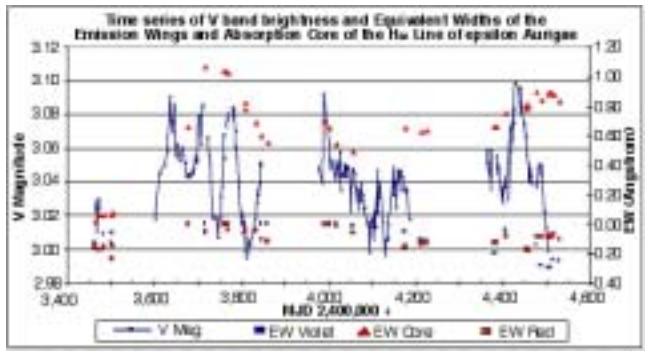

**Figure 9** Correlation of photometric brightness and the equivalent widths of absorption and emission wings of Hα line.

Since JD 2,454,400 the Doppler shifts of the absorption cores of Hα lines have been measured. The Doppler shifts (radial velocity) have increased in parallel with the brightness (V), see Figure 10. Whether this is random coincidence or a real relationship will have to wait further observations.

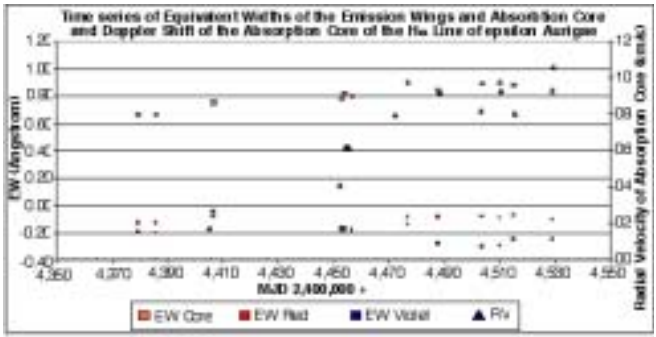

**Figure 10** Correlation of Doppler shift of the absorption core (RV) of Hα line and the equivalent widths of absorption and emission wings.

### Additional Observations

In addition to optical photometry and spectrometry of epsilon Aurigae, observational work in the infrared and ultraviolet has begun in anticipation of the eclipse. Infrared efforts include measurement by Clemens et. al. (2007)[3] of the near infrared spectrum, which shows a mixture of hydrogen absorption and emission lines in January 2006. Mid-infrared spectra were obtained with the Spitzer Space Telescope (Stencel, 2007)[4], that show Fe II emission lines superposed on the F star continuum. As reported at this meeting by Mais and Stencel, interferometric observations were started at the Palomar Testbed Interferometer in order to precisely measure the F star diameter and attempt to detect pulsational changes to it prior to eclipse. Preliminary data reduction suggest the star remains stable with a 2.2 milli-arcsecond diameter, independent of light curve and H-alpha variations. Ultraviolet spectrum observations have been infrequent compared with the previous eclipse, but HST data were reported by Sheffer and Lambert (1997)[5] and Far UV data were reported by Ake (2006)[6] – both reports discuss a model where the emission is due to resonance scattering of photons in the expanding wind of the supergiant or disk from an occulted hot source. Beyond these observations, additional photometry, spectroscopy and polarimetry are still needed in advance of and during eclipse.

### Observing Campaign News

During the 1982-1984 eclipse an Observing Campaign was formed. Newsletters were published and distributed to disseminate the latest information. This was prior to the Internet and while it worked was not high quality and the process was slow. Currently data can be available almost instantly and notices sent to interested parties near instantly.

For the next eclipse we have started a new Campaign. We have 19 people signed up for official participation. These include observers from around the world and around the U.S.A. As of the end of March 2008 we have published five Newsletters. The first one was in September of 2006. Information in these include multiband photometry and spectroscopy as well as information on the observers. These Newsletters are dependent on material received and as we get closer to the eclipse these will be published more often. The current Newsletters are in .pdf format and can be viewed and downloaded on the Campaign web site at:
http://www.hposoft.com/Campaign09.html

We are grateful to the campaign participants and others for sustaining interest in this interesting star system, and encourage everyone to contribute to the Campaign Newsletters with reports of their discoveries by sending email to phxjeff@hposoft.com.

## Conclusion

One of the interesting results to date from the photometry is preliminary evidence for an evolving quasi-period associated with the low amplitude, out of eclipse, light variations. Data from this decade indicate a predominant 66 day period, with other sub periods present (Hopkins and Stencel 2006)[7]. Analysis of data obtained during the decade of the 1990s by Louis Boyd, made available by AAVSO, seems to indicate a 96-97 day period, as mentioned above Post-eclipse observations 1982-1987 by Nha et al.[8], argue for a 96 day periodicity. Even earlier, Shapley[9] reports sparse data over the 1904-11 interval and suggests a 355 day quasi-period, although this cannot be confirmed. Taken at face value, this secular decrease in periods for the low amplitude quasi-variations suggests these will become very fast within a couple of decades, perhaps connected with an accretion death-spiral. Suggestions of changes to periods in epsilon Aurigae were also made by Gyldenkerne[10] who deduced that the 1956 overall eclipse decreased 44 days relative to Gussow's[11] 1936 analysis of 1929 eclipse, while totality was longer by 64 days!

Saito and Kitamura[12] (1986) commented on shortening eclipse phases and propose that the F star is shrinking at a rate of up to 16% eclipse to eclipse. If continued, this effect should be obvious during the coming seasons.

Table 7 shows a summary of best estimates of times of contact during the four eclipses. Data for eclipses prior to the 1982/84 eclipse are assumed to reflect measurements in essentially the V band. We note that duration of totality has apparently INCREASED by ~25% while overall duration and especially egress have been DECREASING. If the trend continues, egress will last only one or two weeks in 2011, whereas ingress will still last 140 or more days! These odd asymmetries suggest dramatic changes on the horizon for epsilon Aurigae in coming decades.

|  | Mean of 2 Eclipses (1902, 1930) | 1955-57 | 1982-84 |
|---|---|---|---|
| Eclipse Duration | 714d | 670d | V 654d |
|  |  |  | B 643d |
|  |  |  | U 630d |
| Totality Duration | 330d | 394d | V 447d |
|  |  |  | B 437d |
|  |  |  | U 455d |
| Ingress Duration | 182d | 135d | V 142d |
|  |  |  | B 135d |
|  |  |  | U 120d |
| Egress Duration | 203d | 141d | V 65d |
|  |  |  | B 71d |
|  |  |  | U 55d |
| Period | 9888d | 9885d | V 9863d |
|  |  |  | B 9875d |
|  |  |  | U 9885d |
| Amplitude | 0.80 m | 0.75 m | V 0.91 m |
|  |  |  | B 0.73 m |
|  |  |  | U 0.84 m |

**Table 7 Summary Timing for the Last Four Eclipses**

### Similar Star Systems

While epsilon Aurigae will not start its eclipse until the summer of 2009 there are three other similar systems that deserve attention.

### EE Cephei

This is an 11 magnitude long period (2,049.53 days/5.6 years) eclipsing binary star system that will begin its 17 day eclipse on 09 January 2009 (2,454,891.30). The system will be in a fairly good position for observing. a good reference for this is by Mikolajewski, et. al.[13].

### Zeta Aurigae

A second long period (972.164 days/2.7 years) eclipsing binary system is zeta Aurigae. Zeta Aurigae will begin its 37 day eclipse on 22 March 2009

(2,454,203.633). While getting low in the Northwest sky, zeta can share the comparison star lambda Aurigae and be imaged in the same frame as epsilon Aurigae using the camera lens CCD photometry. Darling has a good article on this star system in the March 1983 issue of *Astronomy*[14].

**BM Orionis**

A third similar system, and not long period at all, is BM Orionis. BM Orionis or theta 1 Orionis B is one of the four stars in the Great Orion Nebula's (M42) Trapezium. The 6.470525 day eclipse lasts 17 hours and has been likened to a fast eclipse of epsilon Aurigae. This 8th magnitude (V) system is both easy and hard. It is easy to find and theta Orionis D provides a close and good comparison star, but the closeness of the four stars can be a problem as well as the bright background due to the nebula. It presents some interesting challenges, but the eclipse can be observed multiple times from fall through winter. A good paper on this star system is one by Hall and Garrison[15].

**Visual "citizen science" and the IYA**

A wonderful opportunity is emerging with the declaration by the United Nations of 2009 as the International Year of Astronomy (IYA), recognizing the 400th anniversary of the astronomical telescope and Galileo. The US national committee working group on Citizen Science has embraced the eclipse of epsilon Aurigae as a possible centerpiece in its multipronged effort to engage more of the public in skywatching activities. Proposals are under development during 2008, but we hope to bring news and updates to all interested parties via the Campaign Newsletters. One scenario is to promote more visual variable star observing in collaboration with the AAVSO by providing instructional materials to astronomers and astronomy clubs nationally. With these tools, citizens can learn to monitor changing stars like Algol, beta Lyrae and epsilon Aurigae, and report their results to AAVSO and study the cumulative input as part of a larger campaign. In addition, observers with digital imaging equipment can contribute those data as well. In the best case, we expand the number of careful observers who can be in a position to better record the mid-summer eclipse changes in 2010. Some of these observing techniques are summarized above.

**New Book on Epsilon Aurigae**

To support the 2009/11 eclipse and provide background from previous eclipses, two us (Hopkins and Stencel) have written a book on epsilon Aurigae titled *Epsilon Aurigae A Mysterious Star System* that is due to be published in the fall of 2008. For more information and to get a pre-publishing discount, see **http://www.hposoft.com/EAur09/Book.html**